\documentclass[sigconf, authorversion]{acmart}
\usepackage{booktabs}
\usepackage{pgfplots}
\usetikzlibrary{arrows,automata,positioning}
\pgfplotsset{
        compat=1.9,
    }
\AtBeginDocument{%
  }

\copyrightyear{2025}
\acmYear{2025}
\setcopyright{acmlicensed}\acmConference[WWW Companion '25]{Companion
Proceedings of the ACM Web Conference 2025}{April 28-May 2, 2025}{Sydney,
NSW, Australia}
\acmBooktitle{Companion Proceedings of the ACM Web Conference 2025 (WWW
Companion '25), April 28-May 2, 2025, Sydney, NSW, Australia}
\acmDOI{10.1145/3701716.3716885}
\acmISBN{979-8-4007-1331-6/2025/04}

\begin{document}

\title{Bias-Aware Agent: Enhancing Fairness in AI-Driven Knowledge Retrieval}

\author{Karanbir Singh}
\authornote{Karanbir Singh is the corresponding author}
\orcid{0009-0005-2655-6335}
\affiliation{%
  \institution{Salesforce}
  \city{San Francisco}
  \state{CA}
  \country{USA}
}
\email{karanbirsingh@salesforce.com}

\author{William Ngu}
\orcid{0009-0009-0326-0952}
\affiliation{%
  \institution{Salesforce}
  \city{San Francisco}
  \state{CA}
  \country{USA}
}
\email{wngu@salesforce.com}


\begin{abstract}
  Advancements in retrieving accessible information have evolved faster in the last few years compared to the decades since the internet's creation. Search engines, like Google, have been the \#1 way to find relevant data. They have always relied on the user's abilities to find the best information in its billions of links and sources at everybody's fingertips. The advent of large language models (LLMs) has completely transformed the field of information retrieval. The LLMs excel not only at retrieving relevant knowledge but also at summarizing it effectively, making information more accessible and consumable for users. On top of it, the rise of AI Agents has introduced another aspect to information retrieval i.e. dynamic information retrieval which enables the integration of real-time data such as weather forecasts, and financial data with the knowledge base to curate context-aware knowledge. However, despite these advancements the agents remain susceptible to issues of bias and fairness –challenges deeply rooted within the knowledge base and training of LLMs. This study introduces a novel approach to bias-aware knowledge retrieval by leveraging agentic framework and the innovative use of bias detectors as tools to identify and highlight inherent biases in the retrieved content. By empowering users with transparency and awareness, this approach aims to foster more equitable information systems and promote the development of responsible AI.
\end{abstract}




\maketitle

\section{Introduction}
The internet has removed the physical barriers of information access as it delivers limitless knowledge to anyone's fingertips within seconds. With such unending sources of information, efficient information retrieval is a requirement. Information Retrieval (IR) can be defined as the process of obtaining relevant information from a large collection of data based on a user's query. It must be accurate, efficient, and adaptable to ever-evolving technologies and user behaviors, which are the key features of major IR usage today in search engines everywhere \cite{Singhal2001}. The IR process is defined by 2 steps: retrieval of documentation relevant to the user query and ranking the documents by relevancy score to give the most pertinent documents first \cite{Hambarde2023}. In line with these goals, Brin and Page introduce the PageRank algorithm which ranks pages by the number of links and references from other pages, thus effectively ranking them by assigning proper weights based on these criteria \cite{Brin1998}. The efficiency, reliability, and speed of this algorithm in retrieving relevant information is still the backbone of Google's search engine. Building on these foundations, advances in artificial intelligence have paved the way for IR systems that go beyond keyword matching, enabling a paradigm shift toward understanding context and intent. Recently, Large Language Models (LLMs) have disrupted the field of information retrieval. LLMs, such as ChatGPT, excel at retrieving and summarizing knowledge to deliver relevant, accurate, and context-sensitive responses. These advances not only transform communication and democratize knowledge, but also improve information retrieval efficiency, reshaping how we interact and access information in real-world applications \cite{Haque2024} \cite{Zhu2024}.

Although LLMs lead to significant advances in the field of IR, they suffer from problems such as hallucinations and stale information. Because LLMs are trained on static data, the context and data required to answer complex real-world questions where new information arises constantly results in  stale and incomplete data and hence results in hallucinations to fill in the gap with plausible yet factually incorrect answers. Lewis et al. introduce Retrieval Augmented Generation (RAG) to address these challenges. RAG systems consist mainly of two components – Retriever and Generator. The retriever module fetches relevant knowledge from vector stores and a generator that creates context-relevant responses based on parametric knowledge and nonparametric knowledge injected by the retriever into its context window \cite{Lewis2021}. RAG applications were introduced to enhance traditional LLMs by integrating external, up-to-date sources into their responses, improving their ability to handle complex, recent problems \cite{Chen2024}. Building on this, AI agents allow for complex, dynamic, and goal-oriented tasks through dynamic external tool calling, cyclical planning process, and task-specific memory offering many advantages over traditional LLMs and RAGs such as autonomy, decision-making, memory, tool usage, and real-time adaptation \cite{Zhao2023}. Additionally, the ReAct framework extends this concept by having the agent plan its approach to problem-solving through reasoning-evaluating the current task and consideration for possible future steps, similar to logical deduction-and taking action to execute the necessary steps identified in the process \cite{Yao2023}.

Although LLMs' integration into the Information Retrieval system has evolved to its mighty state, there has been an issue with generated information since its first case of documentation: bias. Bias is the systematic imbalance and unfair representation resulting from answers based on sources or assumptions that disproportionately favor or disadvantage groups, often reflecting historical or societal inequities.  Confronting bias is important to not only ensuring fairness and inclusivity but also maintaining reliability and ethical integrity of current and future AI developments. Information, pre-LLM, required user assurance for unbiased, factual sources. After the creation of LLMs, we would only hope that LLMs were fed with curated, fair, and unbiased information to allow everyone's viewpoint to be processed and help adjust the answer given by the LLM. Jaenich et al. found that bias does highly exist inside LLMs, favoring articles like popular, well-known opinions and news sources that garner engagement, which will contribute to unequal representation in the answer generated \cite{Jaenich2024}. Implementing an adaptive reranking system to value by including fairness consequences in the exposure process to equally value under-represented viewpoints. Assuming that an LLM is trained on curated unbiased information and is organized adequately for exposure, as Jaenich et al. expect, the advancements of RAGs are too strong to ignore as they overextend the limits of LLMs. RAGs now contain another issue: even if the original information it is trained on was adequately taken care of, the new information it takes in could also be biased. Wu et al. point out RAGs, when pulling in exterior information, may introduce or even exacerbate fairness issues, and they are heavily reliant on the biases inherent to the external source pulled from \cite{Wu2024}. Since LLMs as reasoners and RAGs as tools are integral components of the agents, they tend to inherently carry bias.

To address these challenges, this paper introduces the Bias-Aware Agent framework\footnote{The source code is available at https://github.com/SinghKaranbir/BiasAwareAgent.}, a system designed to detect bias the content generated by agents. To the best of our knowledge, this is the first attempt to address bias issues using agents. Our contributions are twofold: (1) We propose a modular framework that combines the reasoning capabilities of LLMs with specialized tools for bias detection and retrieval, enabling dynamic and context-aware bias evaluation. (2) We provide a set of queries to showcase the framework's ability to analyze and mitigate biases in real-world scenarios. The queries are provided in Appendix~\ref{appendix:queries}. The rest of the paper is structured as follows: Related Work is discussed in Section 2. Approach is discussed in Section 3. Subsequently, experiments that were conducted to evaluate the approach are provided in Section 4. Finally, conclusion and future work is discussed in Section 5.

\section{Related Work}
In this section, we will discuss the existing work that was done to identify and mitigate bias from AI driven systems. The existing work can be categorized into three types: pre-processing, in-processing, and post processing bias detection and mitigation techniques \cite{Mehrabi2019}. 

\subsection{Pre-processing techniques}
Pre-processing techniques aim to mitigate biases within datasets before they are used for training models, thereby reducing the risk of perpetuating systemic unfairness and thus inherently producing fair models. Kamiran and Calders proposed three data preprocessing techniques: Massaging, Reweighting, and Sampling to address discrimination and mitigate bias in classification tasks \cite{Kamiran2011}. De-Arteaga et al. removed gender-related words from a set of biographies which resulted in significant improvement in the fairness of a classifier used to predict corresponding occupations \cite{DeArteaga2019}.  Raza et al. introduce Dbias, an open-source Python package designed to detect and mitigate biases in news articles. Dbias pipeline is made up of three core modules: bias detection, bias recognition, and de-biasing. The pipeline ensures that pre-processed data is free of bias, resulting in fairer models during training \cite{Raza2022}. 

\subsection{In-processing techniques}
While pre-processing techniques focus on data preparation, in-processing approaches tackle bias directly during model training or inference. The idea is to penalize the model if it favors bias and hence it controls the loss function to minimize bias. For example, Rekabsaz et al. develop AdvBert, a BERT based ranking model that uses adversarial training to simultaneously predict relevance and suppress protected attributes in content retrieved by IR systems \cite{Rekabsaz2021}. Jaenich et al. modify the ranking process using policies to ensure that different document categories are ranked fairly and hence improving fairness metrics by 13\% in IR systems \cite{Jaenich2024}. Singh and Joachims propose a generic fairness-aware learning-to-rank (LTR) framework using a policy-gradient method to enforce fairness constraints within a listwise LTR setting \cite{Singh2019}. Building on this, Zehlike and Castillo integrate fairness into listwise LTR by incorporating a regularization term into the model's utility objective \cite{Zehlike2020}. 

\subsection{Post processing techniques}
Post processing introduces fairness after the model or ranking output is generated. Yang and Stoyanovich proposed fairness measures for ranked outputs and incorporated these measures into an optimization framework to improve fairness while maintaining accuracy \cite{Yang2017}. Zehlike et al. introduced FA*IR, a post-processing algorithm to ensure group fairness in the \ensuremath{top_k} retrieved documents by guaranteeing a minimum proportion of protected candidates while maximizing utility in IR systems \cite{Zehlike2017}. 

In a nutshell, pre-processing ensures unbiased training data, in-processing integrates fairness constraints during model training, and post-processing modifies outputs to achieve equitable results. Our study falls under post-processing, utilizing a classification model for bias detection on content retrieved from a vector store, offering a novel approach to addressing bias in information retrieval systems.
\begin{figure*}[!t]
    \centering
    \includegraphics[width=7in]{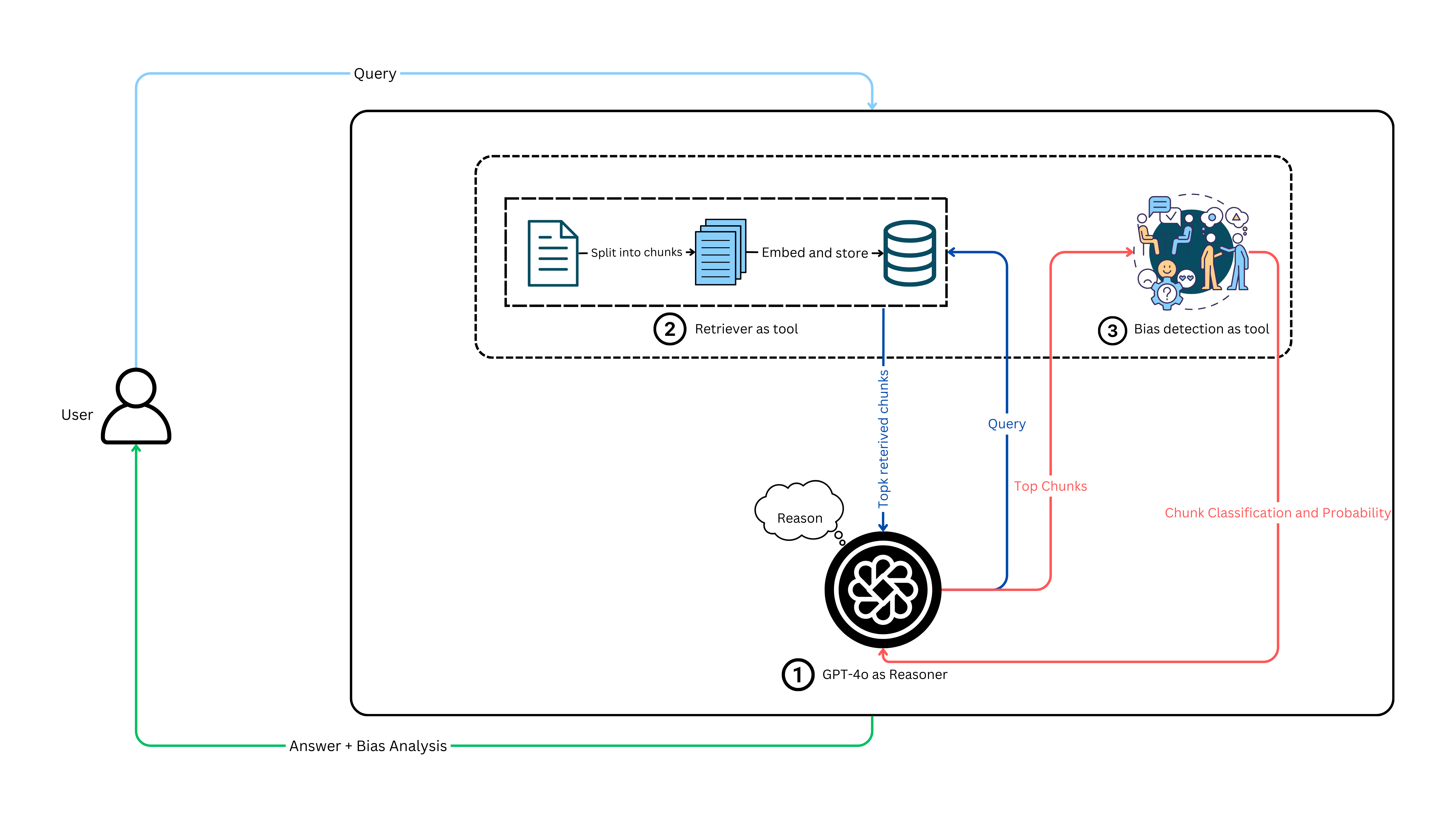}
    \caption{High Level Architecture of Bias Aware Agent. The user submits a query to the agent. 
        The agent processes and reasons on the query, retrieves relevant news chunks from the Vector Store, 
        and ranks them based on relevance. The top-k chunk vectors are passed to the Bias Detection Tool, 
        which analyzes potential biases in the content. Using the output of the Bias Detection Tool, the 
        agent reasons and summarizes the retrieved news while appending a Bias Analysis to the final answer 
        provided to the user. This ensures both relevance and fairness in the response.
    }
    \label{fig:architecture}
\end{figure*}
\section{Approach}
In this section, we explore the internals of the Bias-aware agent framework. We leveraged LangGraph, a robust framework for building agentic systems and developed the agent based on the ReAct agent model. Therefore, we begin by discussing the core principles of the ReAct agent. Finally, we discuss the other components of the framework i.e. focusing on the retriever and bias detection tools as seen in Figure \ref{fig:architecture}.

\subsection{ReAct Agent}
\begin{figure}[!h]
  \begin{tikzpicture}[shorten >=1pt,node distance=5cm,on grid,auto, initial text ={User Query}]
  \tikzstyle{every state}=[fill={rgb:black,1;white,10}]
  \node[state, initial, initial where=above, initial distance=1cm] (s_0) {Reasoner};
  \node[state] (s_1) [text width=2cm,align=center, right of=s_0] {TOOLS (Retriever as Tool, Bias Detection as Tool)};
  \node[state, accepting] (s_2) [text width=2cm, align=center, below of=s_0] {Answer and Bias Analysis};

  \path[->]
  (s_0) edge[bend left, above]               node {Tool call to solve the sub task}  (s_1)
  (s_1) edge[bend left, below]               node {Observation}  (s_0)
  (s_0) edge                                 node {No Tools Needed} (s_2);
  
  \end{tikzpicture}

    \caption{Reasoning loop of Bias Aware Agent which shows how the reasoner can interact with the tools available to the agent and iteratively does the bias analysis over the content retrieved from a vector store.}
    \label{fig:reasoning-loop}
\end{figure}
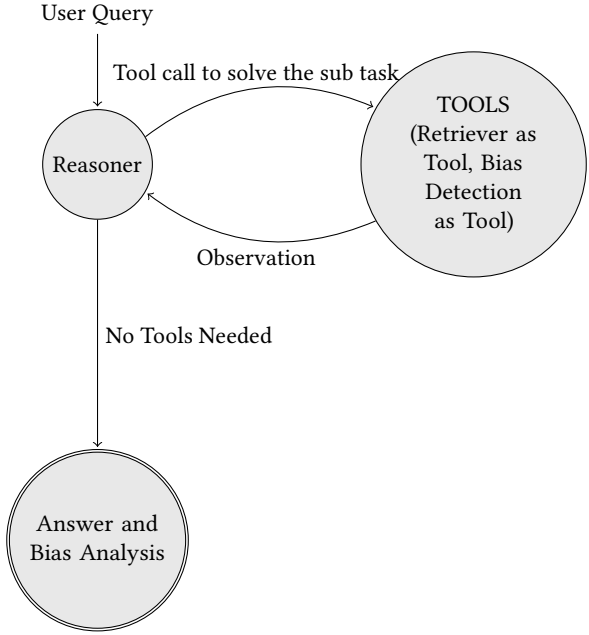

ReAct based agent is designed to solve complex tasks using both reasoning and action capabilities of the LLMs. At its core, it consists of two modules: a reasoner (R) and a set of tools (\ensuremath{\hat{T}}). In our agent, we are using GPT-4o as a reasoner with system prompt shown in Figure \ref{fig:system-prompt}. At each timestep t, the reasoner (R) receives an observation \ensuremath{o_t \in O} from a tool \ensuremath{\hat{t} \in \hat{T}}. It then reasons on \ensuremath{o_t} and based on the current state \ensuremath{s_t \in S} which encapsulates all the context based on previous observations of the agent upto time (t) and updates the state \ensuremath{s_t}  by combining the prior state \ensuremath{s_{t-1}} with the new observation \ensuremath{o_t}. Finally, based on this updated state it then determines an action \ensuremath{a_t \in A}. If the selected action at involves another tool invocation, then the agent invokes the respective tool \ensuremath{\hat{t} \in \hat{T}} and restarts the cycle. Thus, it becomes a reasoning loop which enables the agent to adapt dynamically based on the observations O in the past and effectively addresses complex, multi-step tasks. Figure \ref{fig:reasoning-loop} depicts the reasoning loop of the Bias Aware Agent. 

\begin{figure}[!h]
    \centering
    \begin{tikzpicture}
    % Define the box with rounded corners and orange border
    \node[draw=orange, thick, rounded corners=10pt, inner sep=10pt, text width=8cm, align=left] (box) 
    {
        \textbf{\textless System Prompt\textgreater} \\
        \textbf{You are a highly advanced bias detection system designed to analyze retrieved news articles for bias. Your task is to:}
        
        \vspace{0.3cm}
        \textbf{Answer the query} based on the content of the retrieved article in a concise and factual manner. \\
        \textbf{Analyze} the retrieved content for bias by utilizing available tools. \\
        \textbf{Provide a bias evaluation:} If the output is biased, state: \textit{“This content contains bias.”} \\
        Include a brief explanation of why the content is biased, citing specific examples. Otherwise, output: \textit{“This content appears unbiased.”}
    };

\end{tikzpicture}
    \caption{System prompt for the reasoner of the Bias Aware Agent.}
    \label{fig:system-prompt}
\end{figure}

In the next subsections, we will explore the tools that are available to Bias Aware Agent to analyze the content. 

\subsection{Retriever as tool}
To retrieve relevant content about the user query, the agent utilizes ChromaDB, a vector store as a retriever. The retriever plays a very important role in the working of the agent since it is responsible for fetching relevant documents to analyze the bias on it. 

\subsubsection{Data Ingestion}
To enable effective retrieval and bias analysis, we used news articles since they are known for mixed content (both biased and non-biased) to the database. The process involves first chunking the articles text into small chunks (C) and then subsequently embed them using OpenAIEmbedding model. Now that each chunk \ensuremath{c_i} where \ensuremath{c_i \in C} is transformed into the vector space and can be represented as \ensuremath{v_i}. Finally, \ensuremath{v_i} are inserted into the database.

\subsubsection{Retrieval}
During the reasoning loop, when a reasoner decides to retrieve relevant documents, it generates a query \ensuremath{q} based on the user’s input and current state \ensuremath{s_i}. It then further transforms \ensuremath{q} into its embedding form \ensuremath{v_q} and then similarly search is performed to retrieve \ensuremath{top_k} documents by evaluating distance d between \ensuremath{v_q} and \ensuremath{v_i}. Once the \ensuremath{top_k} documents are retrieved, they are further sent back as observations to the reasoner.

\subsection{Bias Detection as tool}
The Bias Detection tool is essential for identifying and analyzing biases in the content retrieved by the agent. In our implementation, we are using pre-trained text classification model called Dbias which is trained on MBAD dataset to detect bias and fairness in the news articles. This specific model is built on top of distilbert-base-uncased model. \cite{Raza2022} Although, we are utilizing this model as bias detector, but the framework allows the use of any model which can effectively detects bias. Furthermore, inherent bias may exist in the training of any classification model. The framework treats bias detection as a tool, which makes it loosely coupled with the rest of the system. This design ensures that incorporating another version of the model requires minimal changes to the overall application, enhancing adaptability, and enabling continuous improvements in bias detection. 

\subsubsection{Analysis Workflow}
Once the reasoner decides to analyze the retrieved content, it invokes this tool. Bias detector further evaluates the content for the bias and outputs binary classification along with a probability score.

\section{Experimentation}
In this section, we evaluate the bias detectability of Bias-Aware Agent. The section is further divided into three subsections, we first discuss the setup, procedure, and finally findings from the conducted experiments.
\subsection{Experiment Setup}
\subsubsection{Dataset and Queries}
We are using a collection of news articles composed of both biased and unbiased articles as a corpus. As mentioned earlier, we have a tailor-made set of 40 queries: 20 framed to elicit articles with bias and 20 framed to find unbiased articles.
\subsubsection{Evaluation Metrics}
To evaluate the performance of Bias-Aware Agent, we utilize the following metrics: Precision, Recall, F1-Score, and Support. These metrics provide insights into the performance of the agent by examining the classification capabilities of the agent.
\begin{itemize}
\item {\texttt{Precision}}: Precision measures the accuracy of the model when it predicts a sample as belonging to the "positive" class (e.g., biased articles). It quantifies the proportion of true positive predictions among all positive predictions. A higher precision means the model rarely misclassifies unbiased articles as bias sources.
\begin{equation}
    \text{Precision} = \frac{T_P}{T_P + F_P}
    \label{eq:precision}
\end{equation}
\item{\texttt{Recall}}: Recall measures the model's ability to correctly identify all biased articles in the dataset. Equation~\ref{eq:recall} defines the Recall metric. A higher recall indicates the model successfully identifies most of the biased articles, minimizing false negatives.
\begin{equation}
    \text{Recall} = \frac{T_P}{T_P + F_N}
    \label{eq:recall}
\end{equation}
\item{\texttt{F1-Score}}: The F1 score balances precision and recall, combining them into a single value. It is particularly useful when the dataset is imbalanced. The score can be calculated using equation~\ref{eq:f1score}. This score ensures that neither over-prediction nor under-prediction dominates the evaluation.
\begin{equation}
    F\textsubscript{1}\text{-Score} = 2 \times \frac{\text{Precision} \times \text{Recall}}{\text{Precision} + \text{Recall}}
    \label{eq:f1score}
\end{equation}
\item{\texttt{Support}}: Support refers to the number of actual samples in each class (positive or negative), providing context for the other metrics.
Equation~\ref{eq:support-positive} and~\ref{eq:support-negative} refer to the positive and negative support metrics, respectively.
\begin{equation}
    \text{Support} = T_P + F_N
    \label{eq:support-positive}
\end{equation}
\begin{equation}
    \text{Support} = T_N + F_P
    \label{eq:support-negative}
\end{equation}
\end{itemize}

These metrics provide a comprehensive assessment of the agent, ensuring that it can consistently identify bias.
\subsection{Experiment Procedure}
We pass a query to the Agent which will then output the tool logs as well as the AI message. We then parse the logs for the passages that were used, look for the bias classifier and confidence probability. We also save the bias value set to the passage from the original dataset. In the case of multiple articles were used and all aligned to the same bias classifier, it was recorded in our main observation table. On the other hand, if multiple sources were used but had different bias alignments (biased, unbiased, no agreement), we record that separately to simplify our results.
\subsection{Findings}
\begin{figure}[!h]
    \centering
    \includegraphics [width=\columnwidth]{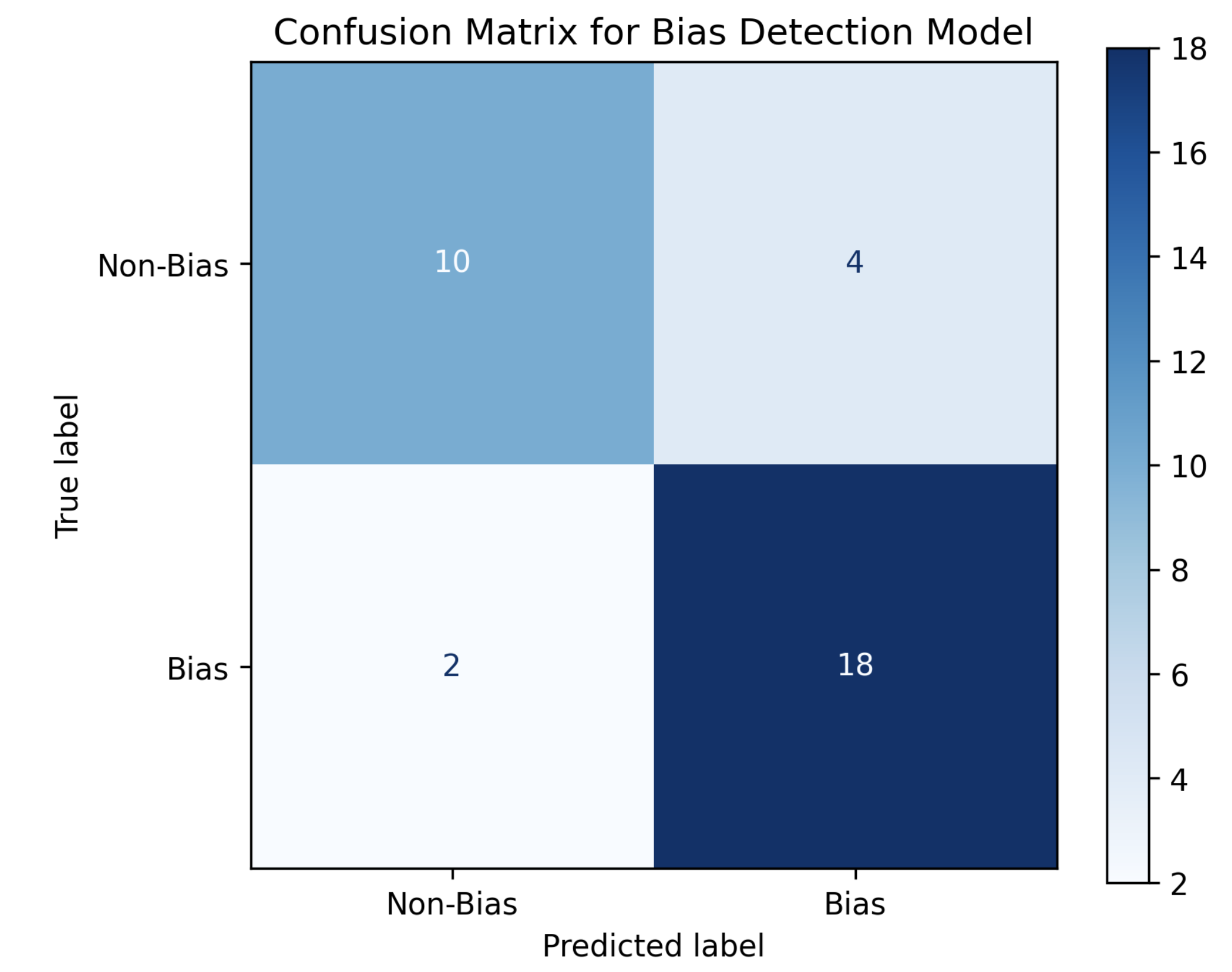}
    \caption{Confusion Matrix showing the performance of the bias detection ability of the agent. Values shown are from queries resulting in analysis of 1 article used a source for a response to a query.}
    \label{fig:confusion_matrix}
\end{figure}

The agent demonstrated a satisfactory bias detection rate, consistently identifying bias and properly associating keyword choice commonly used in biased articles during information retrieval. As shown in Table~\ref{tab:biased_vs_nonbiased}, the weighted average F1-score of 0.795 showcases the agent’s ability to make accurate predictions and  demonstrates a high level of performance. Figure \ref{fig:confusion_matrix} represents the confusion matrix which demonstrates the overall performance of AI agent in terms of answering the query as well as presenting bias analysis.

\begin{table}[h]
    \centering
    \caption{Performance of Bias-aware Agent in detecting bias}
    \label{tab:biased_vs_nonbiased}
    \begin{tabular}{p{1cm} p{1.3cm} p{1.5cm} p{1.5cm} p{1.2cm}} % No vertical lines in ACM tables
        \toprule
        \textbf{Metric} & \textbf{Biased Articles} & \textbf{Non-Biased Articles} & \textbf{Weighted Average} & \textbf{Macro Average} \\ 
        \midrule
        Precision  & 0.818  & 0.714  & 0.773  & 76.60\% \\ 
        Recall     & 0.9  & 0.714  & 0.811  & 80.70\% \\ 
        F1-Score   & 0.857  & 0.714  & 0.795  & 78.69\% \\ 
        Support    & 20   & 14   & -   & - \\ 
        \bottomrule
    \end{tabular}
\end{table}

\begin{table}[h]
    \centering
    \caption{Bias detection performance for individual queries which result in using mixed sources (biased and non biased)}
    \label{tab:performance_with_mixed_sources}
    \begin{tabular}{p{1cm} p{1cm} p{1cm} p{1cm} p{1cm} p{1cm}} % No vertical lines in ACM tables
        \toprule
        \textbf{Query} & \textbf{Biased Articles} & \textbf{Non-Biased Articles} & \textbf{No Agreement} & \textbf{Agent Prediction} & \textbf{Probability} \\ 
        \midrule
        3	&3	&1  &0	&Bias   &0.995 \\
        4	&3	&1	&0	&Bias	&0.593 \\
        5	&1	&2	&1	&Bias	&0.916 \\
        6	&3	&1	&0	&Bias	&0.99  \\
        7	&2	&2	&0	&Non-Bias	&0.745 \\
        27	&3	&0	&1	&Bias	&0.995 \\
        30	&2	&2	&0	&Bias	&0.534 \\
        40	&1	&1	&0	&Bias	&0.569 \\
        \bottomrule
    \end{tabular}
\end{table}

\begin{figure}[!h]
    \centering
    \includegraphics [width=\columnwidth]{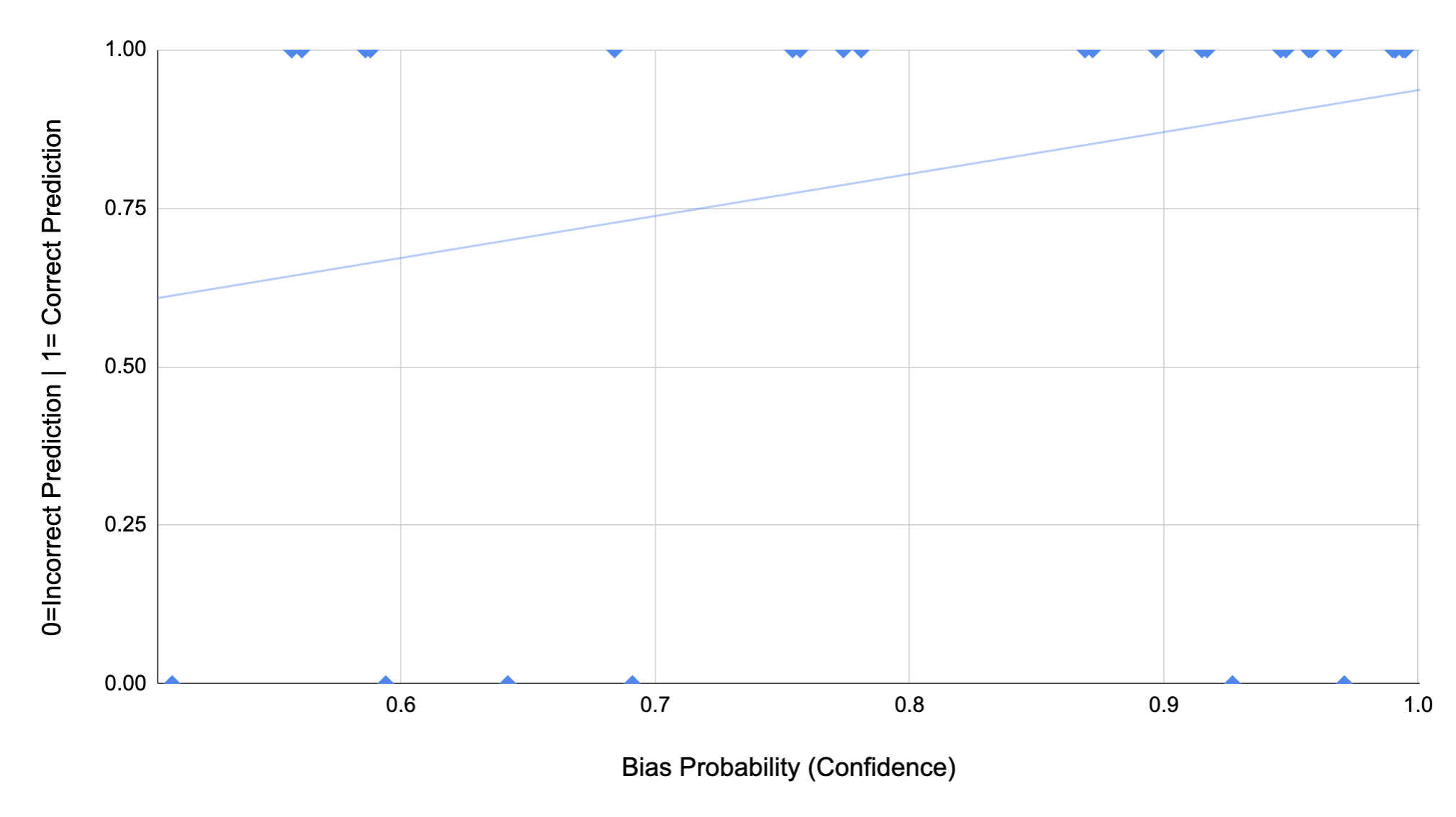}
    \caption{Graph showcasing the correlation between Bias Probability (Confidence) and the correlating Predicted/Actual Result. The Trendline in the graph is auto generated from the data points}
    \label{fig:bias_prob_and_actual}
\end{figure}

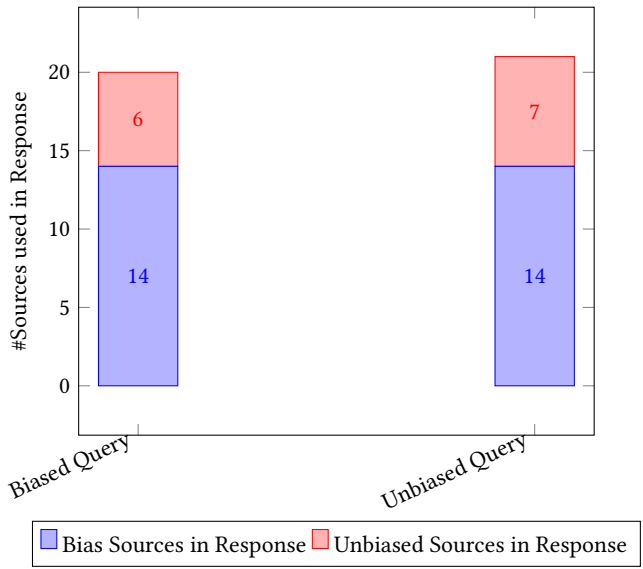
\begin{figure}[!h]
    \centering
    \begin{tikzpicture}
\begin{axis}[
    ybar stacked,
	bar width=30pt,
	nodes near coords,
    enlargelimits=0.15,
    legend style={at={(0.5,-0.20)},
      anchor=north,legend columns=-1},
      ymin=0,
    ylabel={\#Sources used in Response},
    symbolic x coords={Biased Query, Unbiased Query},
    xtick=data,
    x tick label style={rotate=25,anchor=east},
    ]
\addplot+[ybar] plot coordinates {(Biased Query,14)(Unbiased Query,14)};
\addplot+[ybar] plot coordinates {(Biased Query,6)(Unbiased Query,7)}; 
\legend{\strut Bias Sources in Response, \strut Unbiased Sources in Response}
\end{axis}
\end{tikzpicture}
    \caption{Stacked bar chart illustrating the distribution of responses generated from biased and unbiased styled queries. Each bar is divided into two segments: responses where bias was detected in the sources (blue) and non-biased responses (red). The figure highlights the relationship between query formulation and the presence of bias in response content.}
    \label{fig:bar_showing_accuracy}
\end{figure}

Figure \ref{fig:bias_prob_and_actual} shows the data points of the bias tool’s confidence in its analysis to see how well it compares to \ensuremath{T_p} and \ensuremath{F_n}. The y-axis of 1 on the graph represents when the detector accurately depicts the article as biased or unbiased, while a value of 0 signifies an incorrect classification. The average confidence level across all data points is 0.821.  

In the field of Responsible AI, a key challenge lies in balancing utility and fairness. Our approach prioritizes utility by providing analysis so that users can make an informed decision rather than masking bias related terms, enabling users to make informed decisions. The agent’s retriever tool may pull in multiple articles when attempting to gather as much information as possible which can lead to multiple bias levels being tested by an agent’s bias tool which can lead to some mixed results. For the simplicity of this experiment, we documented them in Table~\ref{tab:performance_with_mixed_sources} as outliers and for future improvements to the Agent framework. In comparison to the main query results, it is interesting to see that confidence levels are close to our other average confidence levels of 0.792. It also showcases the effect of using multiple sources to formulate a singular answer and also the effects that 1 or more articles that can be considered ‘biased’ can have a large effect on the decision of biased or non-biased from the agent. The only time a non-biased decision was made was when there were two non-biased sources used alongside two biased sources and even then, resulted in a lower-than-average confidence value for the decision of 0.745.

Figure \ref{fig:bar_showing_accuracy} shows an interesting note that the articles chosen for response generation remained close to even between the Bias and Unbiased styling of the queries. Questions that were created to avoid biased topics or be phrased in a way to attempt to find only facts did not significantly change the amount of bias found in the sources used to gather the data.

\section{Conclusion and future work}
This paper introduces the novel agentic framework which uses bias detection as a tool. It will help users of LLMs to be more knowledgeable and mindful of the responses and information gathered from their questions. Due diligence for users is made easier and allows for the sharing of knowledge to feel safer and more comfortable for users when we know where and how we came to gather the facts. Moreover, this study contributes to advancing transparency in bias-aware flow by highlighting source articles were used and how much bias they carry to generate the overall response of the agent. This transparency enables users to feel more comfortable and have more trust in their usage of agents.

In future work, there are many approaches to furthering the research in this study where it can directly touch on bias mitigation, it could be handled in multiple ways, first bias mitigation as a tool; before the response is generated, a mitigation tool could identify where biases could still exist in its response and perform a part of censoring. Second, there could be a multi-agent system where one agent specializes in removing a specific type of bias from the content in such a way that it can still maintain its utility. Another topic could be advanced source selection and decision-making where an agent can figure out if the bias is detected in the source article and not adding value to the overall response; therefore, it can be omitted. Small improvements to the results or articles retrieval may be accrued from changing to another bias detector or using another retrieval tool. By advancing in any or all of these, knowledge understanding and transparency of information sharing will be an open expansion for all AI outside of this field.

\bibliographystyle{ACM-Reference-Format}
\bibliography{bias_aware_agent}

\appendix
\section{Queries}
\label{appendix:queries}
\subsection{Queries highlighting bias}
\begin{enumerate}
    \item How are abortion providers described in the articles?
    \item What language is used to discuss pro-choice advocates?
    \item How does the article refer to pro-life advocates?
    \item What adjectives are used to describe climate activists like Greta Thunberg?
    \item How are Republican immigration policies framed?
    \item What tone is used when discussing Planned Parenthood?
    \item How are transgender athletes described in the articles?
    \item What narrative is built around anti-vaccine advocates?
    \item How are online platforms regulating or promoting freedom of expression while combating hate speech?
    \item What language is used to discuss activists involved in the Black Lives Matter movement?
    \item How is the Green New Deal described in the articles?
    \item What tone is used in discussing President Trump’s environmental policies?
    \item What are the reported statistics about homelessness rates in urban centers?
    \item What evidence is used to describe the scale and impact of humanitarian crises on affected populations?
    \item How are public officials portrayed in stories about student loan debt?
    \item What language is used to describe Christian leaders or their initiatives?
    \item Why have the trustworthiness of vaccines deteriorate in recent history?
    \item What examples are there of celebrities influencing public perceptions of science and health issues?
    \item What narrative is conveyed around military interventions in the Middle East?
    \item What stance does the article take on transgender rights in sports?
\end{enumerate}
\subsection{Queries highlighting non-bias}
\begin{enumerate}
    \item What are the legislative facts surrounding abortion bans?
    \item How is data on wage disparities across different demographics presented?
    \item What data is used to discuss climate-related disasters?
    \item What are the key details about the benefits and risks of artificial intelligence in healthcare?
    \item How has the increased recommendation of childhood vaccine affect vaccine rates?
    \item What are the reported impacts of late-term abortion laws?
    \item What evidence is cited to evaluate the effects of universal basic income policies?
    \item What are the key facts about the Green New Deal's proposals?
    \item How is economic data tied to immigration policies presented?
    \item What evidence is used to support or refute election fraud claims?
    \item What are the measurable outcomes of COVID-19 policies?
    \item What are some facts about international sports scandals?
    \item What historical information is provided in discussions about voting rights legislation?
    \item How are international efforts to combat cybercrime described?
    \item What are the statistics shared about public health emergencies?
    \item What are the precise details of legislation on gun control?
    \item How are vaccination rates among children reported?
    \item What is the factual basis for claims about unemployment rates among immigrants?
    \item What are the key points about digital media platforms' policies on content moderation and misinformation?
    \item How does the article report on the performance of renewable energy initiatives?
\end{enumerate}
\section{Agent output}
\subsection{Output for query that highlights bias}
\subsubsection{Query}
What tone is used in discussing President Trump’s environmental policies?
\subsubsection{Tool NewsArticleCorpus’s Observation}
The Republican president assumed he was helping the industry at the expense of the environment — a trade-off Trump was happy to make since he rejects climate science anyway.
\subsubsection{Tool Message - bias detector’s Observation}
\["Biased", 0.7811554670333862\]
\subsubsection{Agent Response}
The tone used in discussing President Trump's environmental policies is critical, suggesting that he prioritized industry over the environment and dismissed climate science.

This content contains bias. The passage implies a negative judgment of President Trump's actions by stating he was "happy to make" a trade-off at the expense of the environment and that he "rejects climate science anyway," which suggests a dismissive attitude towards environmental concerns.

\subsection{Output for query that highlights non-bias/facts in articles}
\subsubsection{Query}
What criteria and evidence did the World Health Organization use to declare COVID-19 a public health emergency of international concern?
\subsubsection{Tool NewsArticleCorpus’s Observation}
The head of the World Health Organization said Wednesday its emergency committee needed more information before it was ready to declare the deadly global coronavirus outbreak a public health emergency, even as the hard-hit Chinese city of Wuhan was experiencing severe impact.
\subsubsection{Tool Message - bias detector's Observation}
\["Non-biased", 0.7738906145095825\]
\subsubsection{Agent Response}
The World Health Organization (WHO) declared COVID-19 a public health emergency of international concern after its emergency committee determined that more information was needed to make this decision. The declaration was made in response to the deadly global coronavirus outbreak, particularly as it severely impacted the Chinese city of Wuhan.

This content appears unbiased
\end{document}